\documentclass{article}

\input{tcilatex}

\begin{document}

\title{{\large Just How Final Are Today's Quantum Structures?}\thanks{%
\noindent Contribution to ``Springer Forum: Quantum Structures -- Physical,
Mathematical and Epistemological Problems'', Liptovsky Jan, 1 September 1998.%
}}
\author{{\normalsize P. Busch} \\
{\small Department of Mathematics, The University of Hull}}
\date{}
\maketitle

\begin{abstract}
\noindent I present a selection of conceptual and mathematical problems in
the foundations of modern physics as they derive from the title question.
\end{abstract}

\noindent `Quantum theory' is widely believed to be the ultimate framework
for the description and explanation of physical experience. Daring as such a
`finality prognosis' may be, any attempt at its justification will, I
believe, invariably lead to qualifications: there seems to be no consensus
as to how general the formal framework has to be, nor is there any commonly
accepted physical interpretation, including the question whether it is
necessary or even possible to go beyond the minimal, probabilistic
interpretation. Moreover, there are -- and have always been -- some `clouds
on the horizon', that is, some challenging questions concerning the range of
applicability of quantum theory. There is thus good reason to rephrase the
finality belief into a (hopefully) tractable programme: to determine those
elements of the theory that can defendably be regarded as necessary parts of
any future theoretical physical framework. This is the foundational route of
exploring the scope of the present quantum theory, and I will list some of
the ensuing specific problems. The other line of determining the limits of
the theory lays, of course, in its application in new areas of experimental
research.

1. \textsl{Hilbert space framework.} The significance of the Hilbert space
structure of quantum theory has been largely clarified through the quantum
logic approach, features of the proposition lattice being exhibited as
necessary, \textsl{apriori }conditions of any empirical scientific language.
However, the existence of incommensurable sets of propositions or of
physical systems with or without superselection rules must be regarded as
contingent facts in this approach. Also, the underlying assumption of ideal
measurements of sharp observables can be challenged as an unfounded
idealisation, so that a revision of the programme could be envisaged that is
based on the more general structure of \textsl{effect algebras}. Finally, a
fundamental physical understanding of the selection of the complex field
seems to be lacking.

2. \textsl{Quantum-Classical Connection. }It is possible to formulate
conditions (such as restrictions of the set of states, or superselection
rules) under which a quantum system will appear to display (approximately)
classical behaviour. But a theoretical explanation of why, and under what
circumstances, such classicality conditions come to be satisfied seems to be
lacking. In other words, the emergence of (approximately?) classical
properties in systems of increasing size or complexity has remained
unexplained, and this fact has led some researchers to question the
universal validity of quantum theory. In my opinion, an approach based on
decoherence and environment-induced superselection seems to depend on the
feasibility of a universal state vector/many minds interpretation of quantum
theory. On such an interpretation, decoherence would guarantee the
consistency of the descriptions of physical events given by different
observers; but due to the global entanglement this would come at the expense
of a relativised notion of objectivity where `facts' would appear to have a
rather transient nature. On a perhaps more (or more naive?) realistic view,
the objective description of such classical behaviour as deterministic phase
space evolution requires that on an appropriate macroscopic scale, conjugate
dynamical variables have to be coexistent and thus jointly measurable, so
that their quantum mechanical representatives must be \textsl{unsharp
observables}. Thus I think that the \textsl{fuzzy probability theory}
recently developed by Bugajski and Gudder may offer a natural unifying
framework for a systematic study of the quantum-classical connection.

3. \textsl{Quantum measurement problem.} This is the question of whether and
how the stochastic evolution of states in a measurement, viewed as a
physical process subject to the laws of quantum theory can be reconciled
with the determinism of the unitary Schr\"{o}dinger dynamics, which does not
allow the entanglement of states corresponding to different outcomes to be
broken up. The fact that pointer observables, as macroscopic, classical
quantities do conform with an objective description implies that their
quantum representations are unsharp observables [cf. 2. above]. Thus a
relaxation of the concept of objectification is needed that is applicable to
unsharp pointers, and the question remains whether `unsharp objectification'
opens a way to solving the measurement problem.

4. \textsl{Quantum theory and relativity.} Special or general relativistic
spacetime is a particular instance of a classical domain which turned out
notoriously hard to unify with quantum structures. The formal problems of
relativistic quantum field theory as well as the fundamental conceptual
difficulties encountered in quantum gravity and quantum cosmology remain the
most challenging fields of investigation for foundational research into
quantum structures.

5. \textsl{Axiomatics of time evolution.} The measurement problem and the
problem of time in quantum cosmology require a review of the assumptions
underlying the unitary evolution law for closed quantum systems. A closed
system is a highly idealised construct, as it becomes more and more
difficult with increasing size of systems to shield them from their
environment. Thus, it appears that the only conceivable `large' closed
system is the whole universe. Are the conditions of the theorems of Wigner
and Kadison on unitary representations of dynamical maps applicable in this
singular case?

6. \textsl{Operational quantum theory and quantum information.} On a more
`down to earth' note, quantum mechanics and measurement theory are presently
finding exciting applications in the newly emerging field of quantum
information processing. The rich structure of quantum state spaces, which
encompass the phenomena of superposition and entanglement, seems to open up
entirely new perspectives for enhancing the capacities of communication
channels as well as the security of codes. This calls for a systematic
development of quantum information theory which takes into account the full
structure of quantum probability theory. I see a particularly striking
observation in the fact that for certain purposes, measurements of unsharp
quantum observables are more efficient than sharp measurements.

7. \textsl{Quantum theory for individual objects.} Is quantum theory
`nothing more' than a statistical theory, or could the referent of a quantum
state be a single object? This question reflects a division of the quantum
physics community into two distinct `cultures', corresponding to the two
options it addresses. The task of justifying an interpretation that goes
beyond a minimal probabilistic interpretation amounts to the task of solving
all the quantum puzzles and paradoxes that have made the foundations of
quantum mechanics such an exciting enterprise throughout the history of this
theory. The needs of quantum cosmology, and also the most recent
developments in the experimentation with single microsystems, certainly
encourage the search for a sound individual interpretation. Perhaps the
answer lies in a fuzzy quantum logic, complemented with an unsharp quantum
ontology adapted to a picture of a fuzzy, or soft micro reality.

\end{document}